\documentclass{aa}
\usepackage{natbib}
\usepackage{psfig}
\usepackage{psfrag}
\usepackage{amssymb}
\usepackage[dvips]{graphicx}
\usepackage{afterpage}

\title{Are granules good tracers of solar surface velocity fields?}

\author{M. Rieutord\inst{1,2}, T. Roudier \inst{3}, H.-G.
Ludwig\inst{4,5}, \AA. Nordlund\inst{6} and R. Stein\inst{7}}

\date{\today}
\offprints{M. Rieutord}

\institute{
Laboratoire d'Astrophysique,
Observatoire Midi-Pyr\'en\'ees, 14 avenue E. Belin, 31400 Toulouse,
France \and Institut Universitaire de France
\and Laboratoire d'Astrophysique, Observatoire Midi-Pyr\'en\'ees, 57
avenue d'Azereix, BP 826, 65008 Tarbes Cedex, France
\and Lund Observatory, Box 43, 22100 Lund, Sweden
\and C.R.A.L, \'Ecole Normale Sup\'erieure,46 all\'ee d'Italie,
69364 Lyon, France
\and Theoretical Astrophysics Center and Astronomical
Observatory/NBIfAFG, Juliane Maries Vej 30, DK-2100 Copenhagen, Denmark
\and Michigan State University, East Lansing, MI 48824, USA
}

\newcommand{\infapp}{\raisebox{-.7ex}{$\stackrel{<}{\,\sim\,}$}}
\newcommand{\vu}{\vec{u}}
\newcommand{\vv}{\vec{v}}
\newcommand{\moy}[1]{\left\langle #1\right\rangle}

\begin{document}

\authorrunning{Rieutord et al.}
\titlerunning{Are granules good tracers of solar surface velocity
fields?}

\abstract{Using a numerical simulation of compressible convection
with radiative transfer mimicking the solar photosphere, we compare the
velocity field derived from granule motions to the actual velocity field
of the plasma. We thus test the idea that granules may be used to trace
large-scale velocity fields at the sun's surface. Our results show that
this is indeed the case provided the scale separation is sufficient. We
thus estimate that neither velocity fields at scales less than 2500~km
nor time evolution at scales shorter than 0.5~hr can be faithfully described
by granules. At larger scales the granular motions correlate linearly
with the underlying fluid motions with a slope of $\infapp 2$ reaching
correlation coefficients up to $\sim0.9$.
\keywords{ Convection  -- Sun: granulation -- Sun: photosphere}
}

\maketitle

\vspace*{-1.0cm}
\section{Introduction }

Since the work of \cite{NS88} granules have been used to trace horizontal
flows at the surface of the sun, namely mesoscale flows and supergranulation
\cite[]{Nov89a,Strous95a,Strous95b,RRMV99,RRMR00}. However, assuming
that granules behave like passive scalars is a rather strong assumption
regarding the nature of the granules: these are dynamical structures
which are far from being passive. Unfortunately, the validity of this
assumption has never been assessed and one relies on the hope that
advection of granules (as intensity structures) is statistically
dominant compared to noise processes like the diffusion of temperature
fluctuations and small-scale motions induced by granules.

To make further progress on this issue, we used a simulation of
convection at the sun's surface to test the tracking properties of
granules.  The simulation provides an observable - a time series of
two dimensional images of the emergent intensity - together with the
underlying three dimensional velocity and temperature fields.  This
allows to test and compare horizontal velocities as measured by
different granule tracking techniques against the actual flow
velocities.  The tracking techniques are presently two, namely LCT,
for Local Correlation Tracking, and CST for Coherent Structures
Tracking. LCT has been developed by November and collaborators
\cite[see][]{NS88} and determines the flows from an optimization of
the correlation between two subimages belonging to successive
images. CST was first proposed by \cite{Strous95a,Strous95b} and
recently developed by \cite{RRMV99} and \cite{RRRD01}; it decomposes
each image of the solar surface into a set of granules whose
trajectories are used to derive the velocity fields. Both methods have
shortcomings whose effects can be quantified by the above comparison.

\begin{figure*}[htp]
\centerline{\includegraphics[width=17cm]{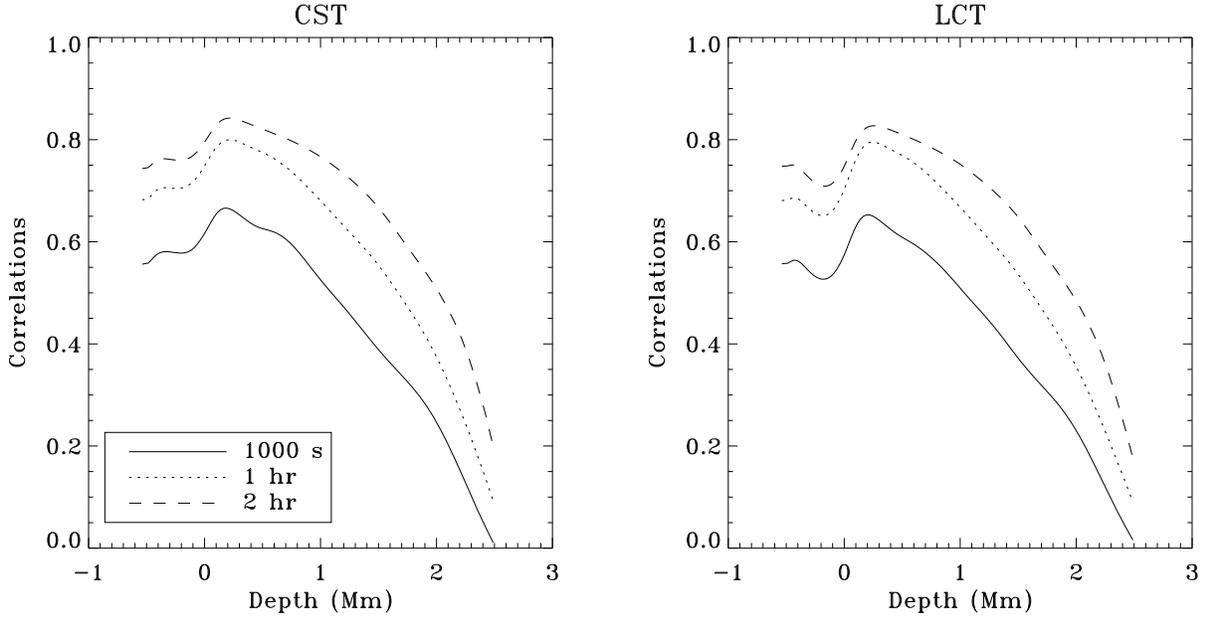}}
\caption[]{Correlations between the velocity field obtained from granule
tracking and the actual velocity field as a function of depth and time-window;
both methods (LCT and CST) give very similar results. A layer 78~km
thick has been used.}
\label{corr1}
\end{figure*}

The simulation used in this letter has been performed using a
compressible convection code coupling fluid motion and radiative
transfer originally developed by two of the authors \cite[for a
description see][]{SN98}.
For this particular run, which aims at simulating supergranulation,
certain trade-offs have been made between physical realism and
computational demands: the radiative transfer was treated in grey
approximation (with frequency independent opacity, with a dependence on
temperature and pressure similar to that of the solar continuum
opacity) and the horizontal resolution was chosen to be a rather coarse
95~km. The restrictions made it affordable to study a large volume
(30$\times$30~Mm$^2$ wide and 3~Mm deep represented by
315$\times$315$\times$82 grid points) which contained several hundred
granules at any given instant in time.

In section 2 of this letter we shall present a global view of the
velocity field, following Euler's viewpoint while in section~3 we
try to characterize the granules in their ability at tracing the flow
field, thus adopting Lagrange's viewpoint. Our conclusion is that
granules are able to trace statistically the large-scale flows but
lead to a systematic underestimation of the actual velocities.

\begin{figure}[htp]
\centerline{ \includegraphics[width=8cm]{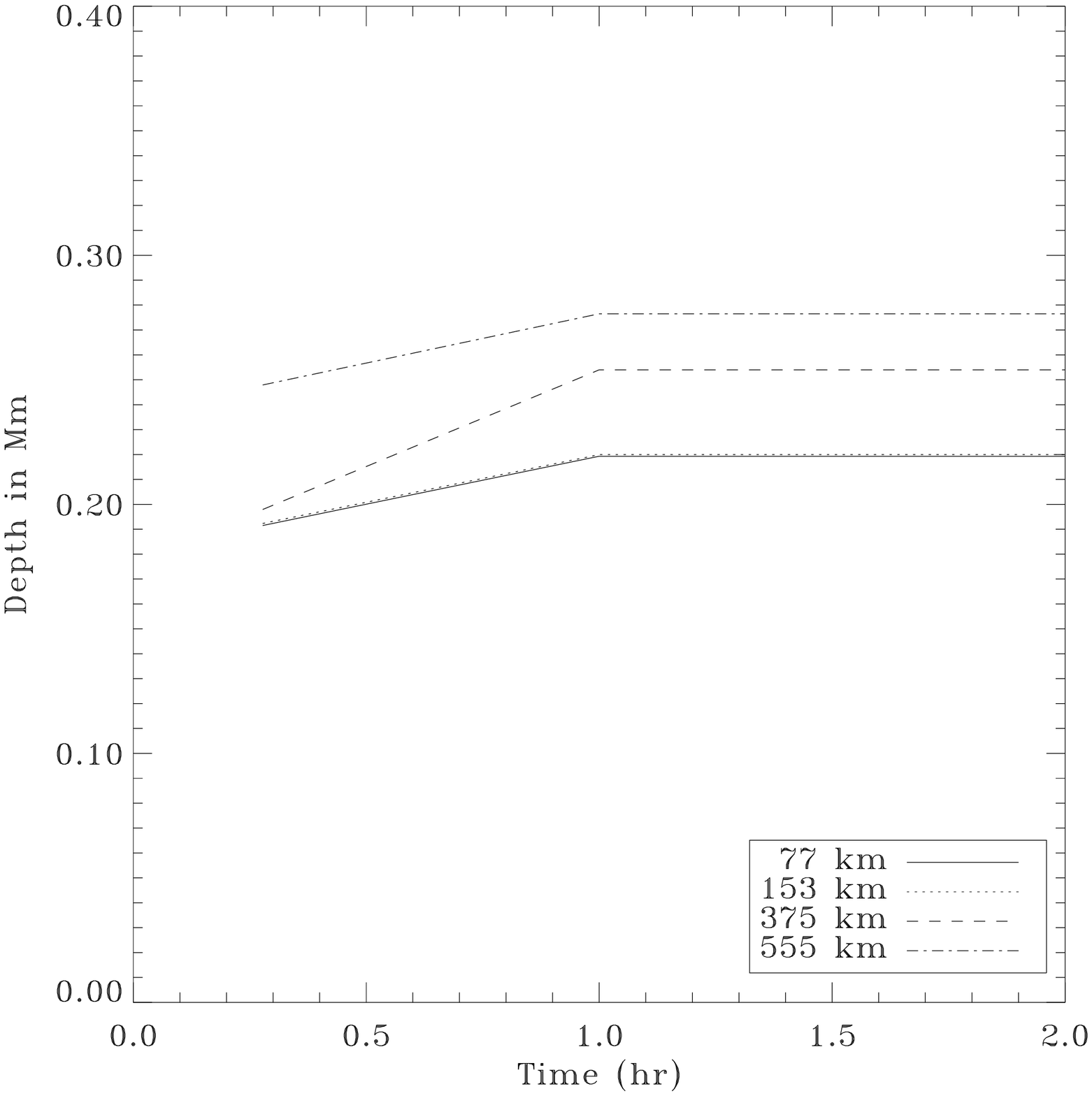}}
\caption[]{Optimal depth for different thicknesses (line type) of the
contributing layer as a function of the time-window. We see that the
depth at which correlation is maximum, remains around 250$\pm$30~km.}
\label{corr2b}
\end{figure}

\section{Euler's view}

\begin{figure}[b]
\centerline{\includegraphics[width=9cm]{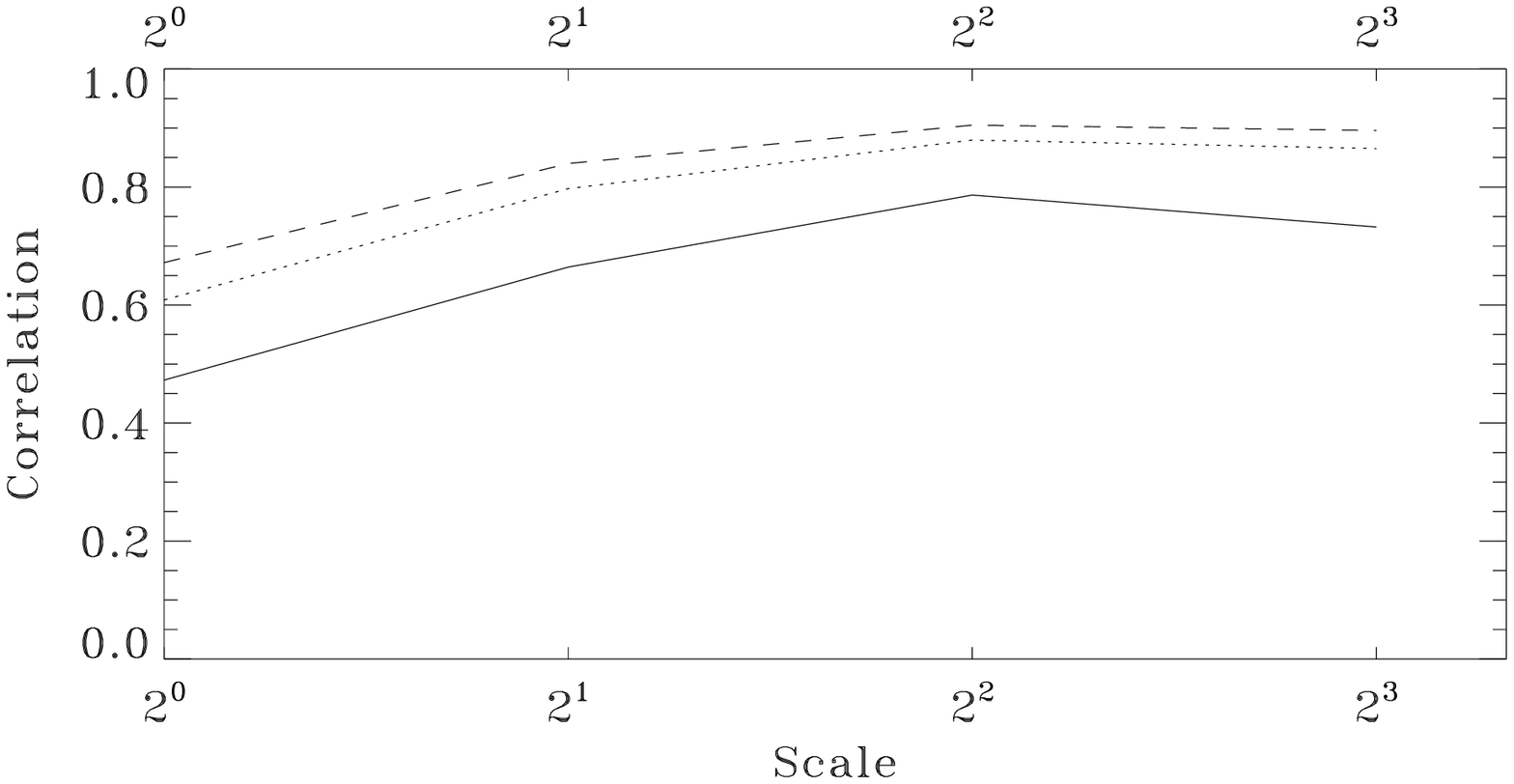}}
\caption[]{Dependence of the correlations with the spatial scale for the
three time averages; the depth used is the optimal depth and the
thickness is 78~km. The $2^0$ scale corresponds to a resolution of
7$\times$7~pixels or 667$\times$667~km$^2$;
we understand the saturation of the $2^3$ scale as a consequence of the
very few resolution elements left for this scale ($\sim 5$$\times$5).
Line type have the same meaning as in Fig.~\ref{corr1}.}
\label{corr_scale}
\end{figure}

As granules are extended test particles, the measured velocity fields
are much less resolved than the one issued from the simulation.
Typically, our granule tracking technique yields a velocity field on a
45$\times$45 grid, i.e.  seven times coarser than the original
315$\times$315 pixels. Hence, for comparison, simulated velocity fields
are rebinned (averaged) to this coarser resolution. Granules also
decrease the time resolution and velocity fields issued from granule
tracking are usually averaged over a time window longer than 5~min; here,
we shall consider three time-windows with durations of 1000~s, 1~h, and 2~h.

\begin{figure*}[t]
\centerline{\includegraphics[width=6cm,angle=90]{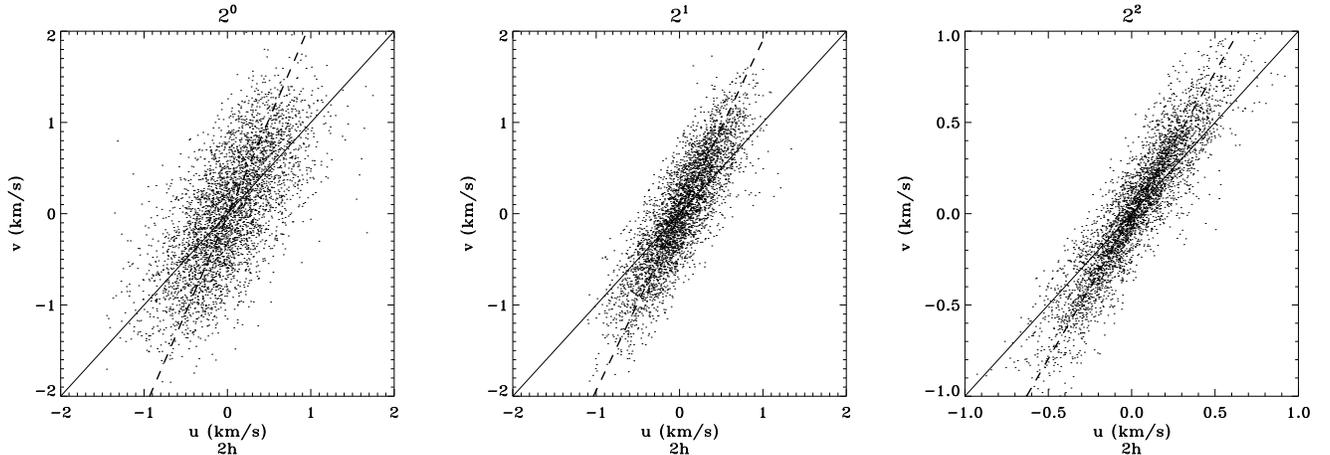}}
\caption[]{For three length scales ($2^0, 2^1, 2^2$), we represent the
actual velocity ($v$ which is either $v_x$ or $v_y$) as a function of
the measured velocity $u$ for a two-hours average. The clouds represent
the $45^2$ grid points. The best fits (dashed line) show that granules
underestimate actual velocities by a factor which varies from 2.1 at
small scales ($2^0$) to 1.6 at large scales (2$^2$).}
\label{corr_cloud}
\end{figure*}

But granules are also three-dimensional structures and therefore they
``feel" the large-scale velocity fields averaged over some range
of depths. The approximate depth and thickness of the contributing
layer need to be determined.

Using the three above mentioned time-windows, we plotted in
Fig.~\ref{corr1} the linear correlation between velocity fields issued from
granule tracking $\vu$ (for a description of how this field is derived,
see Roudier et al. 1999 or Rieutord et al. 2001) and the ``original"
ones $\vv$ as a function of depth; this correlation is defined by
$ C_v = {\moy{\vu\cdot\vv}}/{\sqrt{\moy{\vu^2}\moy{\vv^2}}} $.
This figure clearly shows that the correlation increases with the length
of the time window hence showing that granules are best for tracing
large-scale flows which evolve on long time scales. It also shows that
the correlation is best at a depth below the $\tau=1$ surface (here z=0);
this is clearly emphasized by Fig.~\ref{corr2b} where we see that the
optimal depth (where the correlation is maximum) is between 200~km and
300~km.

We also tested the dependence of correlations with respect to the
thickness of the layer and found that it is weak: variations of
correlations are of 2 or 3\% when the thickness of the layer is varied
between 40 and 900~km.

The foregoing results show that granules do trace long time-averaged
flows, thus we should observe a better correlation when small
spatial scales are filtered out. This is indeed the case, as shown by
Fig.~\ref{corr_scale}. Using the decomposition of the velocity field
onto the different scales yielded by a MultiResolution Analysis with
Daubechies' wavelets, using the scaling function $\phi_4$\cite[]{Daub92},
we show that the correlation reaches $\sim0.9$ at the largest scale
available.

In Fig.~\ref{corr_cloud}, we plotted the actual velocity as a
function of the measured velocity for various length scales. The clouds
of points clearly show that granules motions statistically underestimate
the actual plasma velocity by a factor which (likely) tends to unity as
the scale increases. When no filtering is made, measured velocities
miss ``real" velocities by roughly a factor 2. This quantitative
disagreement is of course even more pronounced in derivative quantities
(divergence or vorticity).

This behaviour is of course no surprise because granules are far from
being passive lagrangian tracers: on the contrary they are active
vortical structures which can move in the background fluid thanks to
their own vorticity or the one of their neighbours. Their motion may
be compared (but just qualitatively) to the random motion of molecules
in a gas: only long time averages or large-scale averages are able to
raise the signal of the mean motion above the noise of random motions.

Finally, let us mention that we have done these tests using the two
presently known methods of granule tracking, namely LCT and CST. As
illustrated in Fig.~\ref{corr1}, both methods give remarkably
close results (within a few percent in correlation), a fact which gives
confidence in the robustness of the results. CST, however, offers additional
informations on the way individual granules follow the background flow;
namely, we can appreciate which granules are the most faithful tracers
and characterize them by some property (size or lifetime for instance). We
discuss this issue in the next section.

\begin{figure}[htp]
\vspace*{-1.7cm}
\centerline{\includegraphics[width=10cm]{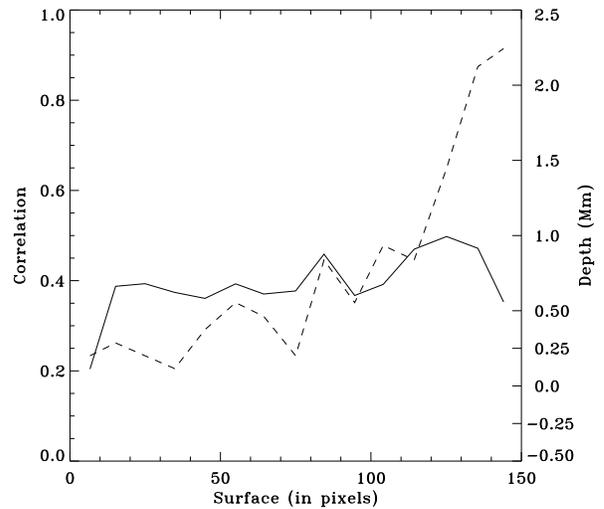}}
\vspace*{-1.5cm}
\caption[]{Maximum correlation (solid line) between the actual
velocity field and the granule displacement at the place of the
granule versus the surface area of the granule. Large granules are
only slightly more correlated than smaller ones. The dashed line
indicates the depth at which correlation is optimum.}
\label{corr3}
\end{figure}

\section{Lagrange's view}

\begin{figure}[htp]
\vspace*{-1.9cm}
\centerline{\includegraphics[width=10cm]{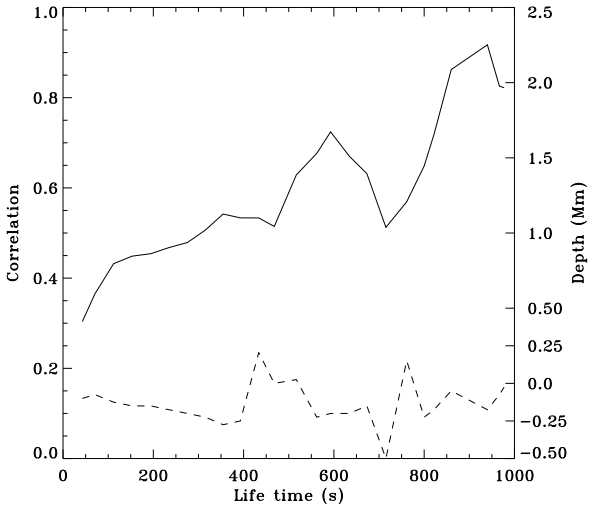}}
\vspace*{-1.5cm}
\caption[]{Maximum correlation (solid line) between the actual
velocity field and the granule displacement at the place of the
granule versus the lifetime of the granule. The dashed line indicates
the depth at which correlation is optimum.}
\label{corr4}
\end{figure}

To see which granules are good or bad lagrangian tracers we computed
the correlation between the mean velocity of individual granules (i.e. the
velocities issued from granule tracking which yield the 45$\times$45
dataset) and the actual velocity at the place of the granule. We
did this computation for different granular sizes (Fig.~\ref{corr3})
and lifetimes (Fig.~\ref{corr4}).

Fig.~\ref{corr3} clearly shows that the size is a poor criterion for
selecting granules whose motions represent the plasma flow. This
figure, however, shows that large granules are sensitive to 'deep'
undercurrents. The depth for optimal correlation increases with the size
of the granules. 

On the other hand, the life-time is a good criterion for sampling the
plasma velocity field. The motion of long-lived granules can reach 0.9
correlation with the actual flow field as shown in Fig.~\ref{corr4}.
Besides, the layer sampled by these granules is not precisely defined
and oscillate around the $\tau=1$ level.

\section{Conclusions}

We have used a simulation of compressible convection with radiative
transfer in grey approximation to test the ability of granules at tracing
the actual plasma flow. The box used for this simulation is
30$\times$30~Mm$^2$ wide and 3~Mm deep, resolved by a 315$\times$315$\times$82
grid. The results of these tests show that
\begin{itemize}
\item Granules tend to be lagrangian tracers when the time and length
scales of the flow tend to infinity: it shows that scale separation is a
necessary condition for using granules at representing plasma flows.
Quantitatively, we find that the length scale needs to be larger than
2.5~Mm and the time scale longer than 1h for the correlation to be
higher than 0.9.

\item They underestimate the velocity field, all the more that scale
separation is weak.

\item Statistically, they probe a layer 300-400~km beneath the $\tau=1$
surface.

\item Long-lived granules are good tracers.
\end{itemize}

Hence, we see that granules can be used as tracers to reveal flows at
meso- and supergranular scale. At smaller scales their own velocity
field has a too strong interaction with the background velocity field.
However, it may well be that velocity fields associated with exploding
granules or with `strong positive divergences' \cite[see][]{RRMR00},
which are near the lowest (allowed) scale, are correctly represented by
granules motions, at least qualitatively, since they are strong advective
motions in nature.

Finally, let us note that the situation in the real sun may not be better
than that of the simulation for the Reynolds number is much higher:
nonlinear interactions are indeed stronger
and thus real granules are less passive. On the other hand,
the drift of the thermal structure with respect to pure advection, which
is another pitfall of granule tracking, is likely correctly represented
by the simulation since it depends on the P\'eclet number which is
modeled accurately by the simulation in the layers of interest.

\begin{acknowledgements}
Calculations have been carried out on the CalMip machine of the
CICT which is gratefully acknowledged. R.F.~Stein acknowledges
financial support by NASA grant NAG 5-9563 and NSF AST grant 98-19799.
{\AA}N acknowledges support by the Danish Research Foundation, through
its establishment of the Theoretical Astrophysics Center.
\end{acknowledgements}

\bibliography{}
\end{document}